# Structural domains associated with permeation, gating and selectivity of Aquaporin ion channels found across phyla.

Andrea J. Yool *
School of Biomedicine and The Institute for Photonics and Advanced Sensing, University of Adelaide, Adelaide SA 5005 AUSTRALIA

* corresponding author: Prof Andrea Yool
email: andrea.yool@adelaide.edu.au
ORCID ID 0000-0003-1283-585x

**Contents**

*Acknowledgement:*

*This is a preprint of the following chapter: Yool, A.J., Structural Domains Associated with Permeation, Gating, and Selectivity of Aquaporin Ion Channels Found Across Phyla, published in Decoding Ion Channels Structure and Function, edited by Avia Rosenhouse-Dantsker, 2026, Advances in Experimental Medicine and Biology 1497: 307-325, reproduced with permission of Springer Nature. The final authenticated version is available online at: http://dx.doi.org/10.1007/978-3-032-07523-9_12.*

***Abstract***
Aquaporins (AQPs) are an ancient group of channels that arose early in phylogeny and radiated throughout the kingdoms of life. Classified as members of the broad Membrane Intrinsic Protein family, AQPs originally were envisioned as strictly water pores that were constitutively open. International research interest is driving new views of AQPs as multi-functional channels with complex control mechanisms that enable adaptive responses to physiological challenges. Expanding numbers of AQP classes are being discovered to allow permeation of diverse solutes in parallel to the archetypal substrate water, with channel activities that show subtype-specific regulation by intracellular and extracellular signals. To date, at least fifteen classes of AQPs from mammals, plants, insects and algae have been proposed to carry ions, enabling amino acid sequence comparisons that suggest ion channel functional domains arise from conserved transmembrane, loop and teminal regions of the tetrameric AQP protein, serving as selectivity filters, gates, and sites for modulation. More AQP ion channel classes likely await discovery, pending identification of relevant activators. Subtype-specific control of dual water-and-ion AQP classes orchestrate diverse roles for AQPs across the domains of life, promoting homeostasis, cell motility, nutrient acquisition, redox protection, sensory detection, and intercellular signaling in cells and tissues.

### *1. Discovery of AQP ion channels*

In groundbreaking work which launched the modern era of aquaporin (AQP) research, expression of cloned AQP1 in Xenopus oocytes was demonstrated to increase the membrane osmotic water permeability, thus identifying AQP1 as a water channel; however, it was reported to not allow transmembrane passage of ions [(1)] based on two-electrode voltage clamp recordings of eight AQP1-expressing oocytes [(2)] in the absence of any activating stimulus (not yet identified at that time). Against this doctrine, AQP ion channels faced a rocky start as a controversial idea. Embedding the concept that AQPs were not capable of ion channel function slowed progress in the field for years. Potential dual ion channel candidates remaining in the vast family of AQPs merit objective consideration; key approaches will be to find native stimuli as needed to activate ion channel gating, and to use an expression system such as yeast in which even small levels of ion conductance conferred by an introduced AQP ion channel can be readily detected by the rescue of growth in a transport-defective yeast line [(3)].

Examples of diverse AQPs endowed with ion channel functionality are being discovered to exist across the domains of life (illustrated in **Figure 1**). The first aquaporin identified as an ion channel was mammalian lens Major Intrinsic Protein (MIP) channel in 1985, followed by soybean Nodulin-26 in 1994, human AQP1 in 1996, mammalian AQP6 in 1999, Drosophila Big Brain in 2002, and subsequently an array of plant and algal PIP (Plasma membrane Intrinsic Protein) and TIP (Tonoplast Intrinsic Protein) channels (**Fig. 1A**). Mammalian AQP1 has been one of the most extensively studied AQP ion channels. In this rapidly emerging field, the available levels of detail for structure-function relationships in ion channel aquaporins vary by subtype; nonetheless, an accumulating set of examples provides a timely tool to assess subtype-specific specializations and evidence for a paradigm shift in our understanding of AQP channel properties.

Aquaporins initially were envisioned as transmembrane pathways just for water. In 1993, soon after the cloned aquaporin gene (*AQP1*) was demonstrated to encode a water channel [(2)], Peter Agre and colleagues suggested "*(T)he term aquaporin should not be used to describe proteins permeated by ions ... or other molecules*" [(4)]. In their Letter to the Editor, they recommended exclusion of MIP (now known as AQP0) because of its permeability to ions [(5)], and the bacterial GlpF (Glycerol Facilitator) based on its permeability to glycerol [(6)]. Nonetheless, the name "aquaporin" was widely adopted for most members of the extensive family, justified by high amino acid sequence homologies and the hallmark presence of pairs of conserved signature motifs (each consisting of asparagine, proline, and typically alanine (NPA)) in all four subunits of the tetrameric channels [7, 8]. Two-electrode voltage clamp recordings were carried out on AQP0-expressing oocytes, with AQP1-expressing oocytes presumably included for comparison as pure water channels. Results of that study showed ion currents were associated with both AQP0 and AQP1 (induced by undefined stretch-activated signaling pathways) that exceeded control oocytes in current amplitude; however these results were interpreted as artifacts that confirmed neither AQP0 nor AQP1 had ion channel functionality [9]. Unexpectedly, one year later the archetypal water channel itself, human AQP1, turned out to be a gated ion channel [10], though the direct triggering stimulus remained to be defined as cyclic GMP in follow-on studies [11].

Transmembrane topology of AQP1 was first evaluated using a chymotrypsin-sensitive epitope introduced by site-directed mutagenesis into a series of positions in AQP1; changes in water channel activity for mutant constructs expressed in oocytes showed that Loop C (linking the 3rd and 4th transmembrane domains, M3 and M4 respectively) was extracellular, and the N- and C- termini and Loop D (linking M4 and M5) were intracellular [12]. Crystal structures have since provided striking three-dimensional views of the AQP channel architecture that supported the original findings, and added substantial detail. Loop B (linking M2 to M3) and Loop E (linking M5 to M6) contain the NPA motifs that line the intrasubunit water pores. Loop D between M4 and M5 is a gating domain [13]. The central passage in AQP1 is lined by transmembrane domains M2 and M5, capped on the outer side by a four-fold symmetry of hydrophobic residues (Val50 in human AQP1) that impose a limiting diameter of less than 3 Å [14] in the closed state.

The predicted structure of the first cloned member of the MIP family, lens MIP (now AQP0), with six transmembrane domains and intracellular N- and C-termini was viewed as consistent with likely functionality as an membrane channel [15]. Basic protein properties of AQPs including tetrameric subunit organization have been noted as a theme shared by many classes of ion channels including potassium channels [16, 17] and others. AQP0 is natively expressed in the lens fibers of the eye where inherited autosomal dominant mutations result in cataracts, thought to be caused by trapping of wild type AQP0 by mutant subunits via oligomerization in the endoplasmic reticulum, inducing cell toxicity [18]. AQP0 protein purified from eye lens was shown to have water channel activity that was blocked by $Ca^{2+}$ (10 μM) when reconstituted in combination with calmodulin [19] which interacts with the carboxyl terminal domain [20]. In the oocyte expression system, AQP0-mediated water permeability was confirmed, though unitary water flux was determined to be approximately 40-fold less than that of AQP1 channels [21]. Further evidence for AQP0 water channel functionality was provided by the loss of water flux after deletion of the C-terminal domain [21].

Ion permeation through purified lens AQP0 was discovered in 1985 by James Hall and colleagues; their work showed reconstituted AQP0 ion channels in bilayers had four conductance states, with a main open state at 200 pS in 100 mM saline that was not blocked by mM $Ca^{2+}$ [5]. Additional work showed bovine AQP0 ion channels in bilayers had two main conductance states of 380 and 160 pS plus subconductance states in 100 mM KCl, were more permeable to anions than cations ($P_{Cl}/P_K \sim 1.8$), and were not sensitive to $Ca^{2+}$ [22]. Interestingly, Ehring and colleagues noted AQP0 channels occasionally closed directly from the maximally open single channel state to a non-conducting state equivalent to bilayer alone (in <1% of all closing events), though intermediate steps were more common. Chicken lens AQP0 (MIP28) in lipid bilayers showed similarly high unitary conductance values (290 pS and 60 pS) in 150 mM KCl saline, and a preference for anions over cations seen by a 5 mV shift in reversal potential when KCl concentration was raised from 100 to 200 mM, indicating a permeability ratio ($P_{Cl}/P_K$) of 1.87 [23].

Most classes of AQP ion channels which have been analyzed at the single channel level, whether reconstituted in bilayers or phospholipid vesicles or recorded in cell excised patches, have typically shown high unitary conductances, long open times, and multiple subconductance states. As seen in AQP0, subconductance states also have been described for AQP ion channels including soybean Nodulin26 and human AQP1 [11, 24, 25]. Nitrogen-fixing soybeans express the AQP Nodulin-26 in the membranes of symbiosomes, prompting the prediction that Nodulin-26 had a transport function; this was confirmed when the purified protein was reconstituted into liposomes and assessed in planar lipid bilayers. Single channel events showed a high conductance (3.1 nS in 1 M KCl saline), and multiple subconductance states. Both anions and cations were permeable, with a weak preference for anions giving a $P_{Cl}/P_K$ ratio of 1.2 based on the reversal potential shift seen when KCl concentration was changed from 1 to 0.1 M [25].

In contrast to the anionic selectivity of AQP0 and Nodulin-26, AQP1 ion channels were found to carry cations [10]. Human AQP1 ion channels were first characterized in voltage clamp recordings in the oocyte expression system, demonstrating cation selectivity and activation by intracellular second messenger signalling [10]. Additional lines of evidence compiled over almost three decades have confirmed that AQP1 channels are directly activated by binding of intracellular cyclic GMP and that ions permeate through the gated central pore, shown using electrophysiology, site-directed mutagenesis, optical probes for cation flux, pharmacology and molecular dynamics modeling [11, 13, 26, 27].

AQP1 in excised patches from oocytes showed a unitary conductance of 150 pS and subconductance states in symmetrical 100 mM $K^+$ salines with $Cl^-$ or gluconate [11]. When reconstituted in proteoliposomes, human AQP1 in symmetrical 112 mM KCl saline showed a single-channel conductance of approximately 75 pS, with subconductance states and patterns of activation that implied possible cooperativity between multiple channels in a patch [24]. Control of extracellular and intracellular gates in tetrameric AQP0 channels also has been proposed to show cooperativity between subunits, modulated by CaM binding [28]. Independent work using recombinant human AQP1 channels confirmed they were activated by cGMP, blocked by a cGMP-related antagonist, and permeable to $Na^+$ and $K^+$ [29] – all properties in agreement with published work [11]; however the reconstituted AQP1 showed much smaller unitary conductances (<10 pS) and a lower proportion of active channels than reported for oocyte assays. Differences in unitary conductances for AQPs between preparations could be due to differences in membrane lipid composition [30], posttranslational modifications including phosphorylation state at the time of harvest [27, 31], interactions with other proteins [32-36], or additional environmental factors.

### *2. Proposed model for ion versus water permeable states*

The structural correlates of the subconductance states observed in AQP ion channels have not been determined, but as a starting point, a working hypothesis can be suggested (**Fig. 1B**), building on a diagram sketched previously [37]. The scheme shows a simple model comprising two opposing stable states-- one favoring water flux via intrasubunit pores (left), and one favoring ion passage through the central pore (right), linked by a series of intermediate states depicting stepwise conformational changes in subunits that might precede central pore opening. A visual metaphor for the outcome is a paper fortune teller (inset top right), though real changes in channel conformation likely involve subtle shifts in barrier residues, not massive reorganization of whole transmembrane domains. The 5th position with central barriers removed but the central pore not fully open could encompass a number of ion current subconductance states, not illustrated. The ability of a channel to conduct ions without all four subunits being in an ion-permissive orientation seems unlikely, given that hydration across the full pore length appears necessary for ion conduction, based on simulation studies [13]. Support for the bimodal idea comes from work showing ion and water conducting functions appear to be reciprocally gated in one of the plant AQP classes (AtPIP2;1), in which ion and water permeabilities were shown to be inversely governed by phosphorylation [31]. Apparently reciprocal gating of water and ion channel activity for AQP1 also was observed in response to phosphorylation of consensus sites by protein kinase C [38]. Validity and details of the hypothetical state model in Fig 1B remain to be determined, but could frame testable hypotheses for further work.

A reasonable prediction of the reciprocal gating idea illustrated in Fig. 1B is that a decrease in water channel activity should occur when AQP ion channels are opened. However, it is important to note this outcome will depend on the proportion of AQP channels in a population that are active as ion channels, which appears to differ between subtypes. While findings for the plant AQP channel AtPIP2:1 support the model [31], unpublished data for AQP1 (DN Birdsell; University of Arizona dissertation research, 1995) showed no appreciable reduction in oocyte swelling rates after AQP1 channel activation by cGMP. Similarly, a mutant of the barley AQP HvPIP2;8 carrying a phosphomimetic substitution in the C-terminal domain did show reciprocal regulation of water and ion fluxes, but the wild type HvPIP2;8 did not [39]. How can this puzzle be reconciled? Taking AQP1 as an example, simple calculations show that the typical microamp currents measured in AQP1-expressing oocytes actually are generated by a tiny proportion (approximately 0.002%) of the total population of AQP1 channels present in the oocyte plasma membrane. The proportion of active ion channels can be calculated directly as the total membrane conductance (typically μS) divided by the unitary conductance (150 pS), as a simplified estimate that does not account for subconductance states. Similarly, the estimated water channel number is determined from the whole oocyte swelling rate divided by the calculated unitary water flux rate (which is the whole cell swelling rate divided by the number of membrane protein channels estimated from freeze-fracture scanning electron microscopic images). In the final analysis, cGMP activation of AQP1-expressing oocytes rendered only 1/56,000 of the channels conductive; nonetheless, this 0.002% level of AQP1 ion channel contribution when assessed in a quantitive model of kidney proximal tubule function translated into cation flux that was comparable to the level seen in native peritubular membrane [17].

Weinstein's model-based calculations further predicted that cGMP stimulation of just 0.002% of AQP1 channels in proximal tubule would increase overall epithelial $Na^+$ reabsorption by 18%, and slightly depolarize the membrane potential by several mV, positing reasonable magnitudes that would be consistent with a realistic response [(17)]. The availability of human AQP1 to be activated as ion channels by cGMP was shown to be controlled by phosphorylation at Tyr 253 in the C-terminal domain [(27)]. Tyrosine dephosphorylation by pharmacological treatment or mutation of Tyr 253 to Cys removed the Tyr phosphorylation signal (confirmed by Western blot) and prevented ion channel activation by cGMP; in contrast, blocking serine and threonine phosphorylation consensus sites by pharmacology or mutagenesis did not alter cGMP-dependent activation responses [(27)]. This example illustrates the principle that when channels are highly abundant in a membrane, only a tiny fraction need to be active as ion channels to confer physiologically meaningful outcomes. As aptly summarized by Bertil Hille, a respected leader in the ion channel field, "Signaling requires only small ion fluxes" [(40)]. The logical corollary is that tissues which depend on high expression levels of an AQP dual water-and-ion channel for water transport must tightly control the parallel ion channel capability in order to prevent collapse of membrane potential and ion gradients.

When coupled with molecular dynamics simulations, crystal structures have enabled impressive insights into the molecular mechanisms of AQP function, elucidating the five-pore organization of AQP1 with a central pore for cations lined by M2 and M5 [(13)], and four individual water pores in each of the subunits [(41, 42)] as shown in **Fig. 1C**. Molecular dynamics modeling identified four hydrophobic residues (Val50, Leu54, Leu170, Leu174 in human AQP1) positioned as barriers to ion permeation in the closed central pore, with positions 50 and 54 located in four-fold symmetry on the outer half of M2, and positions 170 and 174 in the inner half of M5. Dynamics simulations showed that ligand binding to the Loop D gating domain induced widening of the central passage, allowing hydration followed by cation permeation [(13)] as depicted in **Fig. 1D**. Space-filling flow pathways are illustrated for both the ion and water conducting states. Biological assays confirmed that site-directed mutations in the predicted arginine-rich ligand binding site of Loop D dramatically impaired the rate of activation of the ion conductance [(13, 43)].

Since the first introduction of the paradigm-shifting idea of dual water-and-ion channel AQPs, multiple converging lines of evidence have shown that many classes of AQPs can have ion channel activity, which has opened the field for discovery of new classes of multifunctional AQPs.

### *3. Mechanisms of AQP ion channel gating and modulation*

Precisely tuned filtering mechanisms which enable high rates of throughput and selective permeability of intrasubunit pores to water, glycerol and other neutral solutes have been covered capably by other research groups including those of Eric Beitz [44], Roslyn Bill [45], Graça Soveral [46] and many more; this topic is not revisited in this review. In contrast, ion channel gating in AQPs remains an under-researched area of work. Fortunately, accruing examples provide evidence for intracellular ligand binding, gating domains, phosphorylation sites, extracellular pH and salinity as factors controlling AQP channel activities, here outlined for a cross-section of AQP classes.

Identification of Loop D as a gating domain for AQP1 ion channels was supported by molecular modeling and biological assays (**Figure 2)**. Site-directed mutations in Loop D did not prevent AQP1 water channel activity (confirming that the mutant constructs were expressed and targeted to plasma membrane) but did greatly impact ion channel activation [13, 43]. Mutants in which conserved Arg residues at 159 and 160 were changed to Ala in human AQP1 (yellow highlight, **Fig 2A**) showed substantially impaired ion channel activation responses to cGMP as compared with wild type [13, 43]. AQP4 (set in brackets because it has not yet been identified as an ion channel) has high sequence homology with AQP1, and when phosphorylated at Ser 180 (yellow highlight; a position equivalent to Arg 159 in AQP1) by Protein Kinase C showed decreased water permeability [47, 48], indicating Loop D exerts regulatory effects across AQP subtypes. Substitutions of Loop D residues with prolines which introduce kinks in peptide structures had differential effects on ion conductance responses that depended on position. Proline substitutions at Thr 157, Asp 158, or Asp 160 impaired ion channel activation; mutation of Asp 159 to Pro was tolerated without appreciable consequence, but Gly 165 to Pro had the opposite effect, potentiating the conductance response to cGMP [43]. These results provide further evidence that Loop D conformation influences AQP1 ion channel gating.

Interestingly, amino acid sequence alignment shows that Loop D sequences are not conserved across the various types of ion channel AQPs discovered thus far; these classes appear to carry their own conserved motifs and might be predicted to be activated by different subtype-specific types of stimuli (Fig 2A). Loop D in contrast is very highly conserved across AQP1 orthologs [43] in mammals, reptiles and birds (**Fig 2B**), and thus can be deduced to serve a function that offers a selective advantage;

i.e., natural mutations in this domain during evolution have not been well tolerated. Sequence conservation for AQP1 Loop D argues that it enables an essential channel function in native tissues.

The catalyst in 2006 for discovering that Loop D in AQP1 was the gating domain came from the observation of differences between Loop D orientations in two crystal structures [(13)]. In bovine AQP1 (prepared as thalium-derivatized crystals, imaged using multi-wavelength anomalous diffraction) [(41)], the four Loop D segments (one from each subunit) were folded inward, covering the intracellular face of the central pore, as illustrated in the **Fig 2C** drawing. In contrast, in human AQP1 (prepared as two-dimensional crystals, imaged by helium-cooled electron microscopy) [(14)], the arginine-rich Loop D domains were rotated outward, leaving the cytoplasmic vestibule of the central pore accessible. This bimodal conformation was confirmed in molecular dyamic simulations. Dynamics modeling indicated that binding of cyclic GMP to arginines in Loop D (**Fig 2D**) triggered an outward shift of the gating domains (**Fig 2E**), which in turn reoriented hydrophobic side chains of barrier residues lining the central pore, particularly at the internal constriction in M5 (Fig 1D). Subtle changes in barrier residues were then sufficient to induce hydration of the pore, removing a major obstacle to ion permeation and modeled as enabling transition to the open state [(13)].

Single-channel recordings showed that AQP1 channels were relatively slow to activate (on the scale of minutes) after addition of cGMP to the intracellular face of excised patches [(11)], unlike the fast response times measured in seconds for classic cyclic nucleotide-gated channels involved in sensory perception [(49)]. The long latency in AQP1 may indicate that multiple rate-limiting factors in a cascade of conformational changes ultimately lead to full pore hydration, prolonging the time until ion permeation. Since AQP1 in vivo serves in long term functions such as fluid homestasis, volume control, and cell motility, a slower onset of the ion conductance appears to be satisfactory.  Quadruple mutation of the central pore barrier residues (Val 50, Leu 54, Leu 170, Leu 174) to alanines resulted in an inwardly rectifying current response to cGMP, with a shift in reversal potential of approximately +10 mV consistent with an increased permeability to extracellular tetraethylammonium ($TEA^{+}$) ion [(27)].

A consensus on the structural basis for ion channel function in AQP6 is taking shape after a curious history of divergent viewpoints.  Cloned in 1996 and characterized in oocytes, human AQP6 was originally identified as a water channel that was blocked by the classic AQP inhibitor mercuric ion, and not permeable to urea or glycerol [(50)]. Holm and colleagues in 2004 working with rat AQP6 in oocytes found $^{14}C$-glycerol and $^{14}C$-urea did cross the membrane [(51)]. Rat AQP6 differed in showing little baseline water channel activity, being activated rather than blocked by $Hg^{2+}$, and displaying an anion conductance [(52)].

Normally targeted to intracellular vesicles in the kidney collecting duct, rat AQP6 when expressed in oocytes was routed to plasma membrane, a feature that allowed detection of the anionic conductance after exposure to 100 µM $Hg^{2+}$ or acidic pH (<5.5) [(53)]. Anion substitution shifted reversal potential for rat AQP6, indicating a moderately anionic selectivity. Cations also flowed through the channel though less well than anions, yet interestingly $Na^+$ and $NMDG^+$ (N-methyl-D-glucamine) were equally permeable [(53)] which suggested the AQP6 ion pore when open is large. $Na^+$ entry through rat AQP6 was confirmed by $^{22}Na^+$ uptake [(54)]. Mutation of rat AQP6 Lys72 to Glu did not prevent the ion current, but blunted the reversal potential shift seen in wild type upon saline substitution of gluconate for $Cl^-$. It was initially surmised from the location of Lys72 that anions must flow through the intrasubunit pores [(53)]; however, a closer look at sequence alignments suggests that this residue is positioned at the internal end of M2 adjacent to Loop B, and thus sits between the central pore and intrasubunit pore domains, not diagnostic for either permeation pathway. Patch clamp recordings were published in 2002 by Hazama and colleagues for rat AQP6 channels in patches from Xenopus oocytes; the flickery noisy events showed an estimated conductance of 49 pS in 100 mM NaCl, and were activated by 10 µM $HgCl_2$ (via interaction with Cys155 in M4 and Cys190 in Loop E, as identified by site-directed mutagenesis). Ion substitution of $Cl^-$ by gluconate shifted the reversal potential approximately +10 mV; however, the authors concluded anions and cations were equally permeable. Prior work with voltage clamp recordings had supported the idea that $Cl^-$ was more permeable than $Na^+$ ($P_{Cl}/P_{Na}$ ~3.6) in rat AQP6 [(53)]. The current model for AQP6 as an anion conducting channel shows an intriguing resemblance to AQP1 in that the central pore rather than the intrasubunit pore is viewed as the pathway for ion conduction [(55)], suggesting a coherent picture of general AQP structure-function features is emerging (see Section 4 below).

A third example of different mechanisms involved in AQP ion channel activation comes from Drosophila Big brain (BIB), an insect AQP that was first discovered as a channel-like protein of unknown function encoded by a neurogenic gene [(56)]. BIB functions as a monovalent cation channel but interestingly showed no measureable water channel activity in oocytes expressing wild type or terminal-deleted BIB constructs [57]. The monovalent conductance showed a permeability sequence of $K^+$ > $Na^+$ ≫ $TEA^+$ , and was blocked by divalent cations $Ba^{2+}$ and $Ca^{2+}$ but not $Mg^{2+}$ [56, 58]. BIB is regulated by patterns of tyrosine phosphorylation in the C-terminal domain. BIB ion currents were activated by tyrosine kinase inhibition, and conversely inactivated by tyrosine phosphorylation, consistent with a role for BIB as signaling protein involved in regulation of early nervous system development in the fly [56, 58].

AQP1 channel functions are modulated by phosphorylation. Tyrosine phosphorylation in the C-terminal domain increased the proportion of channels available to be gated as ion channels by cGMP [27]. Consensus sites for other kinases enabled opposite patterns of activation of AQP1 water versus ion channel permeabilities, mediated by protein kinase C (PKC) [38]. In that study, phorbol ester activation of PKC caused phosphorylation of AQP1 at consensus sites Thr 157 (located at the proximal end of Loop D) and Thr 239 (in the C-terminal domain), which increased the water permeability but decreased the amplitude of AQP1 cationic currents; neither response was seen in mutants lacking these Thr sites. In contrast, the AQP1 ion current response to cyclic nucleotides was not affected by deletion of PKC phosphorylation sites, consistent with a separate mechanism of action [38], now known to involve direct cGMP binding at the Loop D domain. Some batches of AQP1-expressing oocytes displayed spontaneous activation of ionic conductance after being pricked by insertion of voltage clamp electrodes; this activation response (similar to the originally reported response to forskolin [10]) was blocked by pretreatment with the non-specific kinase inhibitor H7, whereas cGMP-mediated activation was not altered by H7 treatment [11]. There are a number of possible mechanisms of action for effects of phosphorylation state in controlling AQP1 function, which include stabilizing the ion or the water open state conformations, shifting affinity for cGMP, or adjusting the flexibility of Loop D.

## *4. Ionic selectivity of AQP conductances*

A possible structure-functional basis for AQP ion channel selectivity is outlined in **Figure 3**. AQP1 and BIB are cation-conducting AQPs that allow passage of a broad size range of monovalent cations including $K^+$, $Na^+$, $Cs^+$, $Li^+$, and $TEA^+$ [10, 57, 59]. Ions moving through the AQP central pore are expected to transit with their associated hydration shells in place [13]. The AQP ion pore lacks the pattern of amino acids imposing stacked carbonyl groups which allow ion dehydration, prerequisite for ion passage through narrow pores such as those in highly selective $K^+$ channels [60]. The filters in AQPs are needed only to create a generic preference for cations or anions, and thus are expected to be comparatively simple in design.

Insight into possible selectivity mechanisms comes from considering other non-selective ion channels such as the excitatory nicotinic acetylcholine receptors and serotonin receptors which conduct cations, and the inhibitory $GABA_A$ and glycine receptors which conduct anions [(61)]. In their review, Keramidas and colleagues summarized work showing that in anion-selective ligand gated channels, the residues located at the inner boundary of the M2 transmembrane pore domains have a net positive charge; conversely, negatively charged residues are present at equivalent positions in the cation-selective ligand-gated channels. These rings of charge arrayed in the pentameric channel are accessible to permeating ions. An interesting detail for the nAChR and serotonin cation channels is that a positively charged Lys or Arg is located adjacent to the selectivity filter Glu residue, and viewed as being oriented away from the pore axis [(61)]. By analogy with ligand-gated channels, ion selectivity filters for ion channel AQPs would be predicted to involve clusters or rings of charged residues located near the internal or external vestibules of the central pore, creating a net charge opposite to that of the permeating ions.

The extracellular vestibule of human AQP1 has two negatively charged Asp residues in each of the subunits framing the central pore (**Fig 3A**), and creating an 8-point ring of negative charges at the outer face [(24)], with Asp 48 located near the external edge of the M2 pore lining domain, and Asp 185 near the external edge of M5 pore lining domain. Considering a cation selectivity role for AQP1 D185, it is interesting to note that the equivalent position in human AQP6 channel is His 189, which is one of the pH-sensing residues for this channel, as noted below. His residues acquire positive charge at acidic pH, in keeping with expectations for an appropriate residue in an anion-preferring pH-dependent channel. In the human AQP1 crystal structure PDB ID 1FQY [(14)], chosen here because it was fortuitously captured with Loop D in an open-state-like conformation (Fig 3A, right panel), the position of Asp 48 directly above the two central pore barrier residues Val 50 and Leu 54 suggests a position poised to influence ion permeation. Interestingly, the adjacent positively charged Lys at position 51 in human AQP1 appears to be oriented away from the pore axis (Fig 3A, oval inset), which is reminiscent of a similar design in excitatory ligand gated receptors [(61)], though perhaps just coincidence. Justification of these interpretations awaits further investigation.

Amino acid sequence alignment of the M2 pore and adjacent loop regions of known AQP ion channels offers an inaugural opportunity to scan for differences in patterns of net charges in cation- versus anion-preferring subytpes identified thus far (**Fig 3B**). Though speculative, ideas observed here could merit further testing. The cation-preferrring channels (top rows, HsAQP1 to KnPIP) show a predominant pattern of negative Asp residues in the Loop A domain immediately upstream of the M2 pore region (fitting observed cation selectivity), apart from Drosophila Big Brain and algal KnPIP. Presumably if the theory holds, other extracellular sites for BIB and KnPIP (such as Tyr in BIB and Asp in KnPIP, located near the outer edge of M5) or intracellular vestible regions might provide negatively charged positions, but these remain to be identified. Human AQP4 has not been shown to be an ion channel; however, if an activating stimulus were identified, the pattern here predicts it would be cation selective. The anion-preferring AQP channels (lower rows; HsAQP0 to GmNOD26) show a predominant pattern of positively charged Arg or Lys in Loop A close to M2, similarly in keeping with an anionic channel. The exception is GmNOD26, which has a negatively charged Glu residue, as well as tyrosine residues which via their net negative charge of π electrons in aromatic rings can coordinate cation interactions [62]; no other candidates are evident immediately outside M5. The lack of obvious candidates for charge selectivity sites might be consistent with low calculated ionic selectivity value of Nodulin-26 as compared to other anion AQPs; for Nodulin-26 the $P_{Cl}/P_K$ ratio is 1.2, in contrast to $P_{Cl}/P_K$ 1.8-1.9 in AQP0, and $P_{Cl}/P_{Na}$ ~3.6 in AQP6. Further studies will be required to see if the predictions suggested here by sequence alignments are met by experimental results.

Posthoc re-evaluation of prior work on human AQP1 that was modified by double mutagenesis of Asp 48 and Asp 185 to alanines [24] might offer serendipitous support for the charged ring model of AQP ion selectivity, shown in Fig 3. Though the study by Henderson and colleagues concluded there were no significant differences in ionic conductance amplitudes between the wild type and D48A+D185A contructs expressed in oocytes, current-voltage plots in the same paper appear to show a positive shift in reversal potential from -22 mV in wild type to -14 mV in the Asp-deleted mutant (**Fig 3C**), suggesting a reduction in ionic selectivity. Re-testing these channel constructs and other charge site mutations with appropriate controls will be needed to confirm this proposal.

The anion-conducting channel AQP6 has now been analyzed with molecular dynamics modeling, providing fascinating results that have opened a new view of AQP6 structure-function properties, most notably the discovery that the anion conducting pathway is routed through the central pore [55], not the individual subunits. The proposed pH sensor is a pair of His residues located at the outer side of the M5 domain at 184 and 189 in human AQP6. Protonation of the His residues is proposed to cause opening of the central pore filter, formed by Leu 56, Ile 60 and Asn 63 located in the outer half of M2, allowing hydration of the central pore. Positions L56 and I60 in AQP6 are directly equivalent to the AQP1 central barrier residues V50 and L54; in contrast, the third position N63 in AQP6 is a glycine in AQP1 (Fig 3B). The Asn at position 63 in AQP6 is suggested to be the first site of interaction for $Cl^-$ ions as they transit through the pore [55]. The overall design of AQP6 is strikingly similar to that of AQP1, with an interesting difference; in AQP6 the activating stimulus (acidic pH) comes from the extracellular side and opens the outer M2 barrier residues first, whereas in AQP1 the activating stimulus (cGMP) comes from the cytoplasmic side and opens the inner M5 barrier residues first.

It appears that the field is coalescing on a unified view of AQP ion channels that could launch an exciting new era of discoveries.

## *5. Physiological roles of AQPs as multi-solute transporters*

AQP ion channels in plants and animals have essential roles in enabling growth, survival and adaptation to physiological challenges. Structural features and mechanisms of regulation of AQP ion channel activities are being found to have important consequences for maintaining health and viability, but as a dramatic counterpoint, they also raise concerns when AQP multi-functional channels are co-opted in disease progression.

Roles of AQPs in promoting cell motility and influencing neural cell fate are beneficial for natural processes of nervous system development. Colorfully described as creating "bulldozers", AQP1 channels expressed in the leading edges of chick embryo neural crest cells were shown to be essential for faciliating the timely migration of the patterned waves of cells from their point of origin above the neural tube to their target peripheral destinations [63]. Whether the migrating neural crest cells rely on AQP1 water channel activity alone or utilize additional cation channel functionality has not been explored. As reviewed in 2022, cGMP is involved in controlling development of afferent sensory pathways, for example flipping responses of dorsal root ganglion growth cones to guidance cues such as semaphorin [64], indicating contributions of the AQP1 ion channels to neural crest outgrowth are conceivable as part of an array of cGMP-sensitive pathways. Expression of the Big Brain channel in developing fruit flies is essential for neural cell fate determination by a signalling process of lateral inhibition, in which expression of the neurogenic gene product BIB directs cells into epithelial as opposed to neuronal fates, and *Dmbib* gene deficiency results in lethal overgrowth of the nervous system [56]. BIB channels regulated by C-terminal Tyr phosphorylation state showed no water channel activity, but the observed cation channel function [57] is consistent with records of ~20 mV depolarization of resting membrane potentials measured in grasshopper embryos in cells destined for non-neuronal rather than neuroblast cell fates [65]. AQP1 expression in other highly motile cells, such as endothelial cells in angiogenesis and leukocyte immune cells needed for protection and surveillance, suggests that a broad potential range of AQP ion channel roles in cell motility remain to be explored [66].

While AQP1 ion channel involvement in facilitating cell migration benefits normal physiology, it becomes a problem in the context of cancer invasiveness and metastasis. Better understanding the AQP ion channel roles in cancer pathologies could help identify novel therapeutic approaches. Using a $Li^+$-selective photoswitchable probe SHL (Sabrina Heng Lithium) pre-loaded in living HT29 human colon cancer cells, AQP1 ion channel activity was monitored in real time by confocal imaging, with activity detected as lithium hot spots [59]. The hot spots transitioned slowly between on and off states, and were clustered in the leading edges of the motile cancer cells, in agreement with prior experimental work that linked AQP1 expression to acccelerated the rates of cell migration in a variety of aggressive cancer cell lines [67-69], and implied relevance to mechanisms of metastasis. Both $Li^+$ hot spots and cell migration were blocked by treament with an AQP1 ion channel antagonist AqB011 [70] and by siRNA knockdown of AQP1 [59]. The brain cancer glioblastoma is a devastating disease associated with high risk of death typically within two years post-diagnosis, despite the best treatments available. The medical challenge results from the highly invasive nature of glioblastoma cells, a process that has been linked in experimental work in vitro to the activity of AQP1 ion channels [71, 72] located in the leading edges [73]. Agents such as arylsulfonamide AqB011 which block the AQP1 ion conductance by interaction with the Loop D gating domain [70] show promise in slowing glioblastoma cell migration and could be lead compounds for adjunct treatments aimed at holding the cancer cells in place while prolonging the window of therapeutic opportunity for primary eradication treatments [74].

The first example demonstrating that an aquaporin ionic conductance can have a physiologically relevant function was in brain choroid plexus, in which natively expressed AQP1 channels generated a cGMP-dependent cation conductance that was capable of adjusting net rates of cerebral spinal fluid production [75]. Whole-cell patch electrophysiology of the choroid plexus cells confirmed permeability to $Na^+$, $K^+$, $TEA^+$, and $Cs^+$, voltage insensitivity, and cGMP dependence, with background $K^+$ and $Cl^-$ currents minimized by using Cs-methanesulfonate recording salines. Cation current responses to cGMP in primary cultures of rat choroid plexus were decimated after siRNA knockdown of AQP1 expression, which had no effect on native $K^+$ or $Cl^-$ currents. The AQP1 cation current was activated by stimulation with atrial natriuretic peptide (ANP) which works via endogenous receptor guanylate cyclase receptors; ANP caused a decrease in the rate of basal-to-apical fluid transport, an effect blocked by the AQP1 ion channel antagonist $Cd^{2+}$ which restored normal fluid transport rates [75]. Evidence for AQP1 ion channel function in choroid plexus provides a mechanism consistent with the known effects of ANP in slowing cerebral spinal fluid production in vivo [76]. It is possible that cation entry through AQP1 channels results in local accumulation of $Na^+$ in the internal vestibule region that could reduce osmotically driven fluid efflux through the water pores, essentially putting the brakes on fluid excretion.

AQP1 ion channels in the retinal pigmented epithelium (RPE) of the eye provide a mechanism for fluid absorption from apical-to-basolateral side [77] in a process that helps protect against fluid accumulation in the subretinal space, a factor in retinal detachment diseases [78]. In monolayer cultures of human RPE, apical-to-basolateral fluid transport measured in a filter chamber was stimulated by treatment with a membrane-permeable cGMP agonist, and by receptor-mediated activation of cGMP by ANP; these responses were blocked by the AQP1 antagonist AqB013 but not the kinase inhibitor H-8, consistent with direct binding of cGMP to the target of action [77] in the facilitation of fluid uptake by AQP1. It is possible that AQP1-mediated cation accumulation in the internal vestibule region could augment osmotically driven fluid influx through the water pores, essentially promoting fluid absorption.

AQP1 ion channels in red blood cells appear to be involved in the dynamic regulation of membrane compliance, a property that is essential as these 8 μm diameter cells course through the blood stream and must compress repeatedly, on average once a minute over their four month lifespans, to fit through the narrow diameters of capillaries (~5 μm) in the microvasculature. The AQP1 ion conductance in human red blood cells was discovered to be an unexpected co-target for the therapeutic actions of the furan compound 5-hydroxymethyl furfural and related derivatives in human sickle cell erthrocytes [79]. Protection from the morphological stiffening (sickling) of red blood cells resulted from a reduction in a mysterious cation leak current first described in 1993 as "$P_{sickle}$" [80] and now identified as AQP1, which when blocked protects cells from sickling [79], an effect augmented by the parallel complementary action of furan drugs in reducing mutant hemoglobin aggregation.

Plants use AQP ion channels to adapt to drought stress conditions. A non-selective cation current described as a $Ca^{2+}$- and pH-sensitive pathway for $Na^{+}$ entry into Arabidopsis roots was found in 2002 [81], but the molecular identity of this channel remained unknown for years, until 2017 when AtPIP2;1 was identified as a dual water-and-ion channel aquaporin with the same properties of localization, regulation and function [82], solving the mystery. AtPIP2;1 is regulated by phosphorylation at Ser residues in the C-terminal domain; mutations at the consensus sites to Asp (mimicking phosphorylation) were associated with a high ion conductance and low water permeability, whereas replacement by Ala (mimicking dephosphorylation) favored high water permeation and a low ion conductance [31]. Other plant AQPs including AtPIP2;2 also carry non-selective cation conductances [31, 82, 83]. Maize ZmPIP2;5 is a cation channel permeable to $Li^{+}$, illustrated using the SHL photoswitchable probe for optical monitoring of $Li^{+}$ entry across root cell membranes [84]. Expression levels of the anion-conducting AQP channel OsPIP1;3, when boosted in roots of rice plants in response to drought stress, enhance the capacity for transporting nitrate anions in addition to water (Liu et al., 2020).

In summary, there is an impressive range of functional roles for AQP ion channels being uncovered, including identification of new roles in health and disease, as well as the resolution of old mysteries around leak currents that have eluded molecular identification even though recognized as key components of physiological responses.

## ***6. Conclusions and Outlook***

Aquaporins, once thought to be permeable to water alone, are being acknowledged as having complex repertoires of permeable substrates and diverse arrays of specializations and mechanisms of regulation. Physiological relevance of the AQP ion channels is highlighted by research advances that are defining the ion channel functions and the structural mechanisms which mediate their multifunctional roles.

An adaptive benefit of AQP ion channels through evolution appears evident in the fact that amino acid sequences of ion channel pore and gating domains are highly conserved within given classes of aquaporins across species. AQP ion channels are key contributors to diverse processes well beyond fluid homestasis alone. More classes of aquaporin ion channels are likely to be discovered pending identification of the signaling pathways that govern activation. A unified view of structure-function properties of AQP ion channels is emerging. Further work is anticipated to continue to challenge assumptions, expand the breadth of documented roles for these surprisingly sophisticated proteins, and to spark to new interest in AQP research across all the domains of life that could open a wealth of unexpected discoveries.

***Figure Legends***

**Figure 1. Overview of aquaporin ion channels.** (**A**) Illustration of the variety of organisms in which aquaporins with ion channel function have been discovered thus far. (**B**) Schematic diagram of a simplified state diagram illustrating a hypothetical bimodal system with two stable states, carrying water (left) or conducting ions (right), linked by substates reflecting conformation changes in hydrophobic central pore barrier residues at the subunit level (in random order). (**C**) Images from structural dynamics modeling [13] of the AQP1 tetramer (left) viewed from above showing water and ion pore locations, and (**D**) viewed from the side, showing the positions of barrier residues in human AQP1 central pore. [Panels A and B were created in BioRender. Yool, A. (2025) https://BioRender.com/j14n077. Images in C and D were used with permission from Elsevier, Copyright Clearance Center #501959597, relabeled to match human AQP1.]

**Figure 2. Sequence comparisons and structural features of the Loop D gating domains.** Loop D is located between the 4th and 5th transmembrane domains (M4, M5). (**A**) Multiple amino acid sequence alignment was done with NIH Cobalt (www.ncbi.nlm.nih.gov/tools/cobalt; conservation setting 3 bits), showing highly conserved (red), moderately conserved (blue) and variable regions (gray) that illustrate differences between AQP classes. Yellow highlights in AQP1 show Arg residues needed for cGMP binding, and the Gly residue which potentiates activation after mutation to Pro; yellow highlight in AQP4 shows Ser that when phosphorylated by PKC decreases water permeability. (**B**) Weblogo illustration of the frequency of occurrence of Loop D amino acids as a function of position, compiled for AQP1 channels from mammals, reptiles, and birds, and showing high sequence conservation [43]. (**C**) Cartoon diagram of labeled Loop D residues in the closed configuration. (**D**) Molecular dynamics model-generated image capturing cGMP binding to the arginine-rich region in AQP1 Loop D. (**E**) Model-generated images showing Loop D of AQP1 viewed from the intracellular side in the ion channel closed state (left) and open state (right). [Panels B to E are from Yu et al [13], used with permission, Copyright Clearance Center #501959597, edited to create the cartoon in (C), and relabeled to match human AQP1.]

**Figure 3. Proposed model for the structural basis of ionic selectivity filters in aquaporin ion channels.** (**A**) Images for an AQP1 crystal structure (PDB ID 1FQY) thought to be in an ion channel open-state-like conformation. Left: top view showing positions of two aspartate residues (D48 and D185) surrounding the central pore

(dotted circle). Right: side view of M2 barrier residues Leu 54 and Val 50, with Asp 48 directly above in the permeation pathway. Oval inset: orientation showing the side chain of Lys 51 extending away from the central pore axis. (**B**) Amino acid sequence alignment of extracellular Loop A and the adjoining pore domain M2, showing positions of amino acids with negative (red), positive (blue) charges, and Tyr residues (yellow), for cation-selective AQPs (rows HsAQP1 to KnPIP), AQP4 (not shown to be an ion channel), and anion-selective AQPs (rows HsAQP0 to GmNOD26). (**C**) Current-voltage plots of ion currents recorded in oocytes expressing wild type AQP1 (left) or a double mutant with Asp 48 and Asp 185 replaced by Ala (right) from Henderson et al [(24)]. [Panel A (left side) and Panel C graphs were used with permission from Elsevier, Copyright Clearance Center # 501959601, edited to mark x-axis positions of reversal potentials.]

FIGURE 1

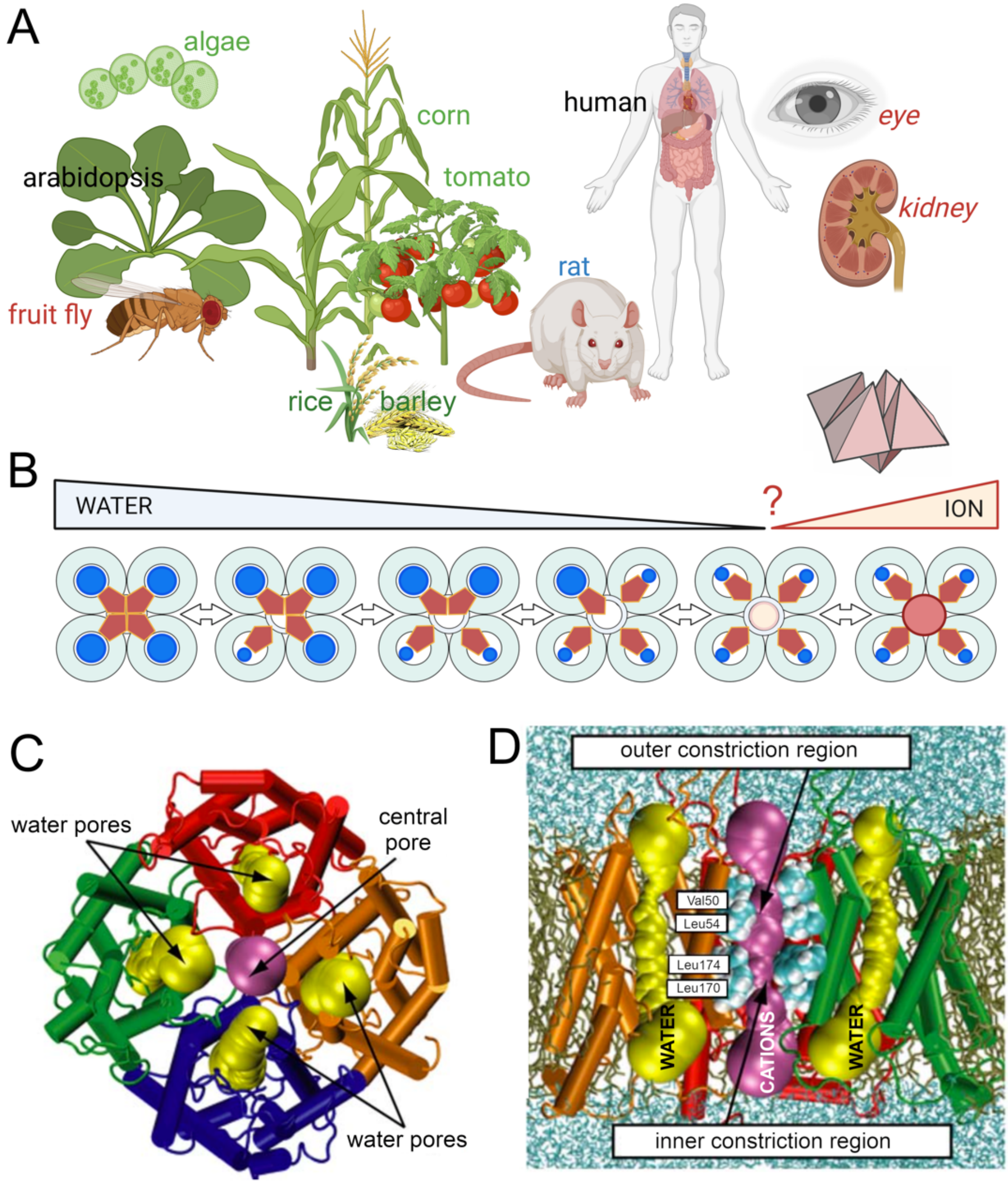
A
algae
arabidopsis
fruit fly
corn
tomato
rice
barley
rat
human
eye
kidney
B
WATER
?
ION
C
water pores
central pore
water pores
D
outer constriction region
Val50
Leu54
Leu174
Leu170
WATER
CATIONS
WATER
inner constriction region

FIGURE 2

A

```
                   M4 |      loop D         | M5
HsAQP1      154   LATT-DRRR------RDLGGSAPLAIGLSV
AtPIP2;1    186   FSAT-DPKRSARD--SHVPVLAPLPIGFAV
AtPIP2;2    184   FSAT-DPKRNARD--SHVPVLAPLPIGFAV
ZmPIP2;4    190   FSAT-DPKRSARD--SHVPVLAPLPIGFAV
ZmPIP2;5    186   FSAT-DPKRNARD--SHVPVLAPLPIGFAV
SnPIP2;5    182   FSAT-DPKRNARD--SHVPVLAPLPIGFAV
HvPIP2;8    193   FSAT-DPKRIARD--PHVPVLAPLLIGFSV
DmBIB       205   FVST-DPMKKFMG-------NSAASIGCAY
KnPIP       232   FAAT-DPKRSALEVSPHLGVLAPLAIGFSI
[HsAQP4]    175   FASC-DSKR------TDVTGSIALAIGFSV

HsAQP0      146   FATY-DERRNGQL------GSVALAVGFSL
HsAQP6      160   FAST-DSRQTS--------GSPATMIGISV
OsPIP1;1    196   FSAT-DAKRNARD--SHVPILAPLPIGFAV
OsPIP1;3    195   FSAT-DAKRNARD--SHVPILAPLPIGFAV
ZmTIP1;1    158   YATAVDPKK------GSLGTIAPIAIGFIV
GmNOD26     172   CGVATDNRA--------VGELAGIAIGSTL
```

B

C

D

E

FIGURE 3

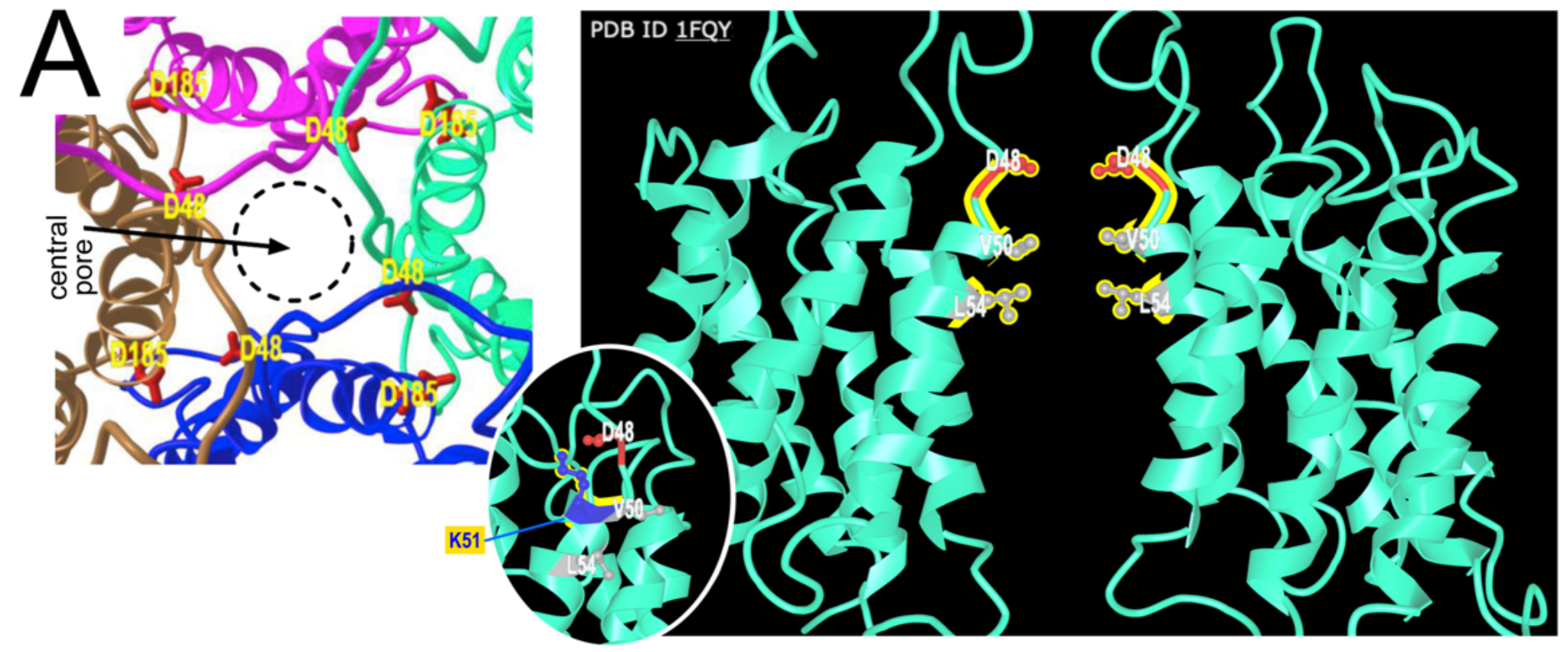

A
PDB ID 1FQY
D185
D48
D185
D48
central pore
D48
D185
D48
D185
D48
V50
L54
D48
V50
L54
D48
K51
V50
L54

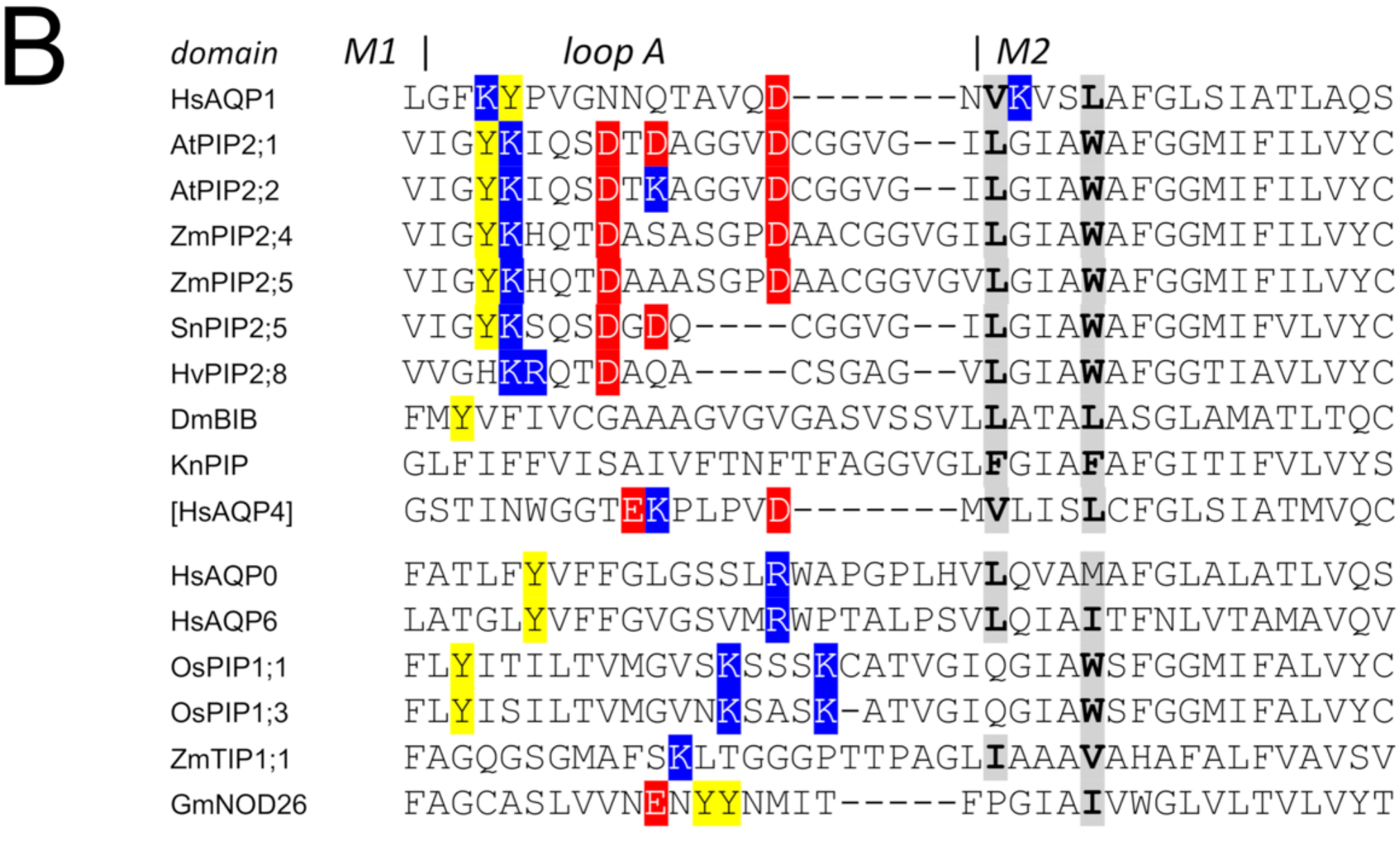

B
domain M1 | loop A | M2
HsAQP1 LGFKYPVGNNQTAVQD-------NVKVSLAFGLSIATLAQS
AtPIP2;1 VIGYKIQSDTDAGGVDCGGVG--ILGIAWAFGGMIFILVYC
AtPIP2;2 VIGYKIQSDTKAGGVDCGGVG--ILGIAWAFGGMIFILVYC
ZmPIP2;4 VIGYKHQTDASASGPDAACGGVGILGIAWAFGGMIFILVYC
ZmPIP2;5 VIGYKHQTDAAASGPDAACGGVGVLGIAWAFGGMIFILVYC
SnPIP2;5 VIGYKSQSDGDQ----CGGVG--ILGIAWAFGGMIFVLVYC
HvPIP2;8 VVGHKRQTDAQA----CSGAG--VLGIAWAFGGTIAVLVYC
DmBIB FMYVFIVCGAAAGVGVGASVSSVLLATALASGLAMATLTQC
KnPIP GLFIFFVISAIVFTNFTFAGGVGLFGIAFAFGITIFVLVYS
[HsAQP4] GSTINWGGTEKPLPVD-------MVLISLCFGLSIATMVQC
HsAQP0 FATLFYVFFGLGSSLRWAPGPLHVLQVAMAFGLALATLVQS
HsAQP6 LATGLYVFFGVGSVMRWPTALPSVLQIAITFNLVTAMAVQV
OsPIP1;1 FLYITILTVMGVSKSSSKCATVGIQGIAWSFGGMIFALVYC
OsPIP1;3 FLYISILTVMGVNKSASK-ATVGIQGIAWSFGGMIFALVYC
ZmTIP1;1 FAGQGSGMAFSKLTGGGPTTPAGLIAAAVAHAFALFVAVSV
GmNOD26 FAGCASLVVNENYYNMIT-----FPGIAIVWGLVLTVLVYT
grey = barrier residues in HsAQP1
V50 L54

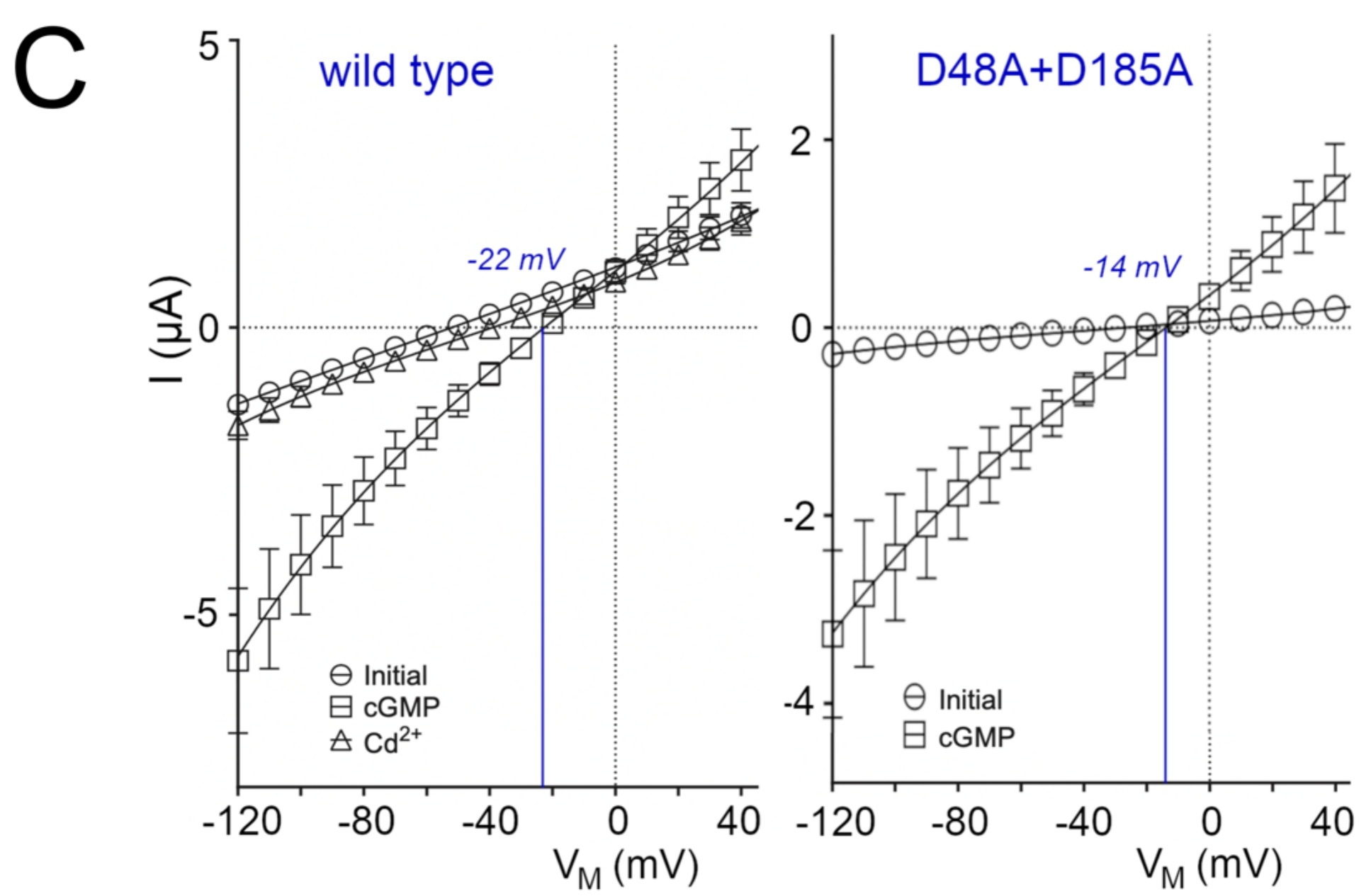

C
wild type
D48A+D185A
I (µA)
5
0
-5
2
0
-2
-4
-22 mV
-14 mV
Initial
cGMP
Cd2+
Initial
cGMP
-120
-80
-40
0
40
VM (mV)